# Molecular causality in the advent of foundation models

## Authors

- **Sebastian Lobentanzer** ✉
  - ⓘ [0000-0003-3399-6695](#) · [slobentanzer](#) · [slobentanzer](#)
  - Heidelberg University, Faculty of Medicine and Heidelberg University Hospital, Institute for Computational Biomedicine, Heidelberg, Germany

- **Pablo Rodriguez-Mier**
  - ⓘ [0000-0002-4938-4418](#) · [pablormier](#)
  - Heidelberg University, Faculty of Medicine and Heidelberg University Hospital, Institute for Computational Biomedicine, Heidelberg, Germany

- **Stefan Bauer**
  - ⓘ [0000-0003-1712-060X](#) · [Vis7i](#)
  - Helmholtz AI and TU Munich, Munich, Germany

- **Julio Saez-Rodriguez** ✉
  - ⓘ [0000-0002-8552-8976](#) · [saezrodriguez](#) · [saezlab](#)
  - Heidelberg University, Faculty of Medicine and Heidelberg University Hospital, Institute for Computational Biomedicine, Heidelberg, Germany

✉ — Correspondence possible via [GitHub Issues](#) or email to Sebastian Lobentanzer <sebastian.lobentanzer@gmail.com>, Julio Saez-Rodriguez <pub.saez@uni-heidelberg.de>.

## Introduction

Correlation is not causation. As simple as this widely agreed-upon statement may seem, scientifically defining causality and using it to drive our modern biomedical research is immensely challenging. Since being described by Aristotle approximately 2500 years ago [1], causal reasoning (CR) remained virtually unchanged until it experienced significant formal and mathematical advancements [2,3,4] and a resurgence in the field of machine learning [5,6] only in recent times. In parallel, biomedicine has made major leaps in the past century, in particular in the development of high-throughput and large-scale methods.

In the field of systems biology, however, great hopes of causal insights from large-scale omics studies have largely been thwarted by the complexity of molecular mechanisms and the inability of existing methods to distinguish between correlation and causation [7,8,9].

[Randomised clinical trials](#) show that, in a lower-dimensional context, we can reliably identify causal effects. By controlling "all" relevant covariates in a trial (via the principle of the gold-standard, randomised, double-blind, and placebo-controlled trial), we isolate the causal effect of the controlled variable, i.e., the treatment. In the language of Pearl's [Do-Calculus](#) [10], we measure the outcome of, for instance, do("Treat with Vemurafenib") when conducting a clinical trial on V600E-positive melanoma [11]. However, translating this mode of reasoning into the high-dimensional space of modern omics poses enormous challenges. The dramatically larger parameter space of models at the molecular level leads to problems in the performance of methods and the identifiability of results [12,13,14], as well as in model explainability [15]. With this perspective, we discuss the current connections between CR and molecular systems biology in the context of these challenges. We will elaborate on three main points:

- biases and what they mean for CR, particularly in the context of biomedical data

- the role of prior knowledge (PK) in CR and how to translate PK into suitable biases

- the role of foundation models in molecular systems biology and their relationship to CR

# Background

## Causal discovery and inference

The field of CR distinguishes between causal discovery - the process of building causal hypotheses from data - and causal inference - the process of predicting specific outcomes given data and the causal relationships known about the system *a priori*.

Causal discovery is more expensive than inference both computationally and data-wise, because it involves distinguishing between correlation and causation and extracting generalisable relationships from the data [12,16]. For modern systems biology, this means that methods for causal discovery typically require large amounts of experiments. Highly parameterised models such as neural networks increase this requirement even further. As such, many regard causal discovery in molecular biomedicine as a scaling problem [17,18].

Causal inference, in contrast, focuses on quantifying the causal effects of one variable on another within the framework of already hypothesised causal relationships. This approach leverages PK about the assumed causal links, which in the causal field are often encoded using directed graphs. Most inference mechanisms perform better when including PK at some point in the process, as has been observed in biomedical research [19]. This allows researchers to represent both the causal connections between variables and their directionality, which is required to understand how changes in one variable might lead to changes in another. For instance, in the case of the RAF-MEK-ERK signalling pathway, a graph would depict RAF activation leading to MEK activation, which in turn leads to ERK activation. This clear representation of directionality is important for causal inference, as it ensures that analyses focus on the effect of upstream changes on downstream outcomes. For example, in analysing phosphoproteomic data to assess the impact of inhibiting RAF, a graph-based approach would guide researchers to correctly attribute subsequent changes in ERK to this specific intervention. Without this causal framework, one might mistakenly interpret correlations as bidirectional influences or overlook confounding factors, leading to incorrect conclusions. However, inference is also very sensitive to the completeness of the PK that is applied; most biomedical PK is far from complete [20]. For instance, the function of more than 95% of all the known phosphorylation events that occur in human cells is currently unknown [21,22]. In contrast to causal discovery, scaling therefore plays a smaller role in causal inference; here, the main problems are incompleteness and identifying the "right" biases to apply.

## The ladder of causality

Orthogonally to the distinction between causal discovery and inference, we can also distinguish between different levels of causality. Pearl's ladder of causality roughly distinguishes three types of CR in increasing order of power: observation, intervention, and counterfactuals [23]. While the inferences we wish to make in biomedical research are often of the counterfactual type (e.g., "would RAF inhibition lead to decrease in ERK activation if the media contained Epidermal Growth Factors?"), the data we have available is typically observational (e.g., "the levels of RAF and MEK activity are correlated") and sometimes interventional (e.g., "targeting RAF with CRISPR leads to a decrease in ERK activity"). To generate interventional or even counterfactual inferences from observational data is a major challenge, if not impossible, depending on the characteristics of the system under study [24].

There are approaches to delineate interventional inference from observational data, such as the 'natural experiments' framework [3,4]. However, these approaches are by their nature even more data-hungry than when using interventional data, as they often do not use the full breadth of the dataset [25]. Therefore, in biomedical research, there has been a push towards generating large-scale interventional data, for instance through the use of CRISPR/Cas9 screens with single-cell resolution [26]. Current developments of CR in the biomedical field thus mostly focus on these types of data.

## Deduction and induction

Lastly, in CR, we can also distinguish between deductive and inductive reasoning. Deductive reasoning is the process of deriving a conclusion from a set of fixed and known premises. "All men are mortal, Socrates is a man, therefore Socrates is mortal" is a classic example of deductive reasoning. In biomedical research, this is typically the process of deriving a conclusion from a set of PK. For instance, having PK of the linear activation cascade EGFR->RAS->RAF->MEK->ERK->Growth, and that Vemurafenib will inhibit RAF activity, allows us to deduce that giving Vemurafenib will inhibit growth of cancer cells [11].

Inductive reasoning, on the other hand, involves making generalisations from specific observations. Testing the hypothesis above, we apply Vemurafenib in a clinical trial of V600E-positive melanoma and find that it is clinically efficacious [11]. Commonly, we then use induction to infer from this limited cohort that the treatment may be effective in the entire population. We could further infer that Vemurafenib may be an effective remedy in other V600E-positive cancers as well, or that inhibiting this cascade may be a general mechanism of action of anti-cancer agents in cancers that display MAPK pathway overactivation [27]. In the molecular realm, we could further infer that the inhibition of other components of the cascade, such as EGFR or MEK, may also be promising target leads [28].

The main difference between deduction and induction is that the former is logically complete - i.e., if the premises are true and the argument is valid, the conclusion must also be true. However, deduction is also more limited in scope than induction. In biomedical research, we often have to rely on inductive reasoning because we cannot feasibly test all hypotheses in a deductive manner. In consequence, the *inductive biases* we introduce into our models (i.e., those mechanisms in the model that help with inductive reasoning) are a pivotal part of performing CR in biomedical research.

# Bias

## Meaning and examples of biases

Biases, generally, are systematic prejudices of a model towards certain outcomes. Humans make frequent use of biases to function in a complex world with limited cognitive resources [29]. In fact, we often presume causality from observation (i.e., we "jump to conclusions"), which is indicative of a strong inductive bias [30]. A good *heuristic* is the application of a suitable bias to a problem, such that the solution can be considered acceptable despite limited resources.

In machine learning, we can distinguish between useful and harmful biases. Harmful biases are common issues in the technical process of training models; they include, for instance, sampling bias, selection bias, and confirmation bias [12,31]. While addressing harmful biases is a crucial part of machine learning, we will not discuss them further in this perspective.

Useful biases, on the other hand, are biases that are introduced into a model to improve its performance. Since most models developed in biomedical research and the broader machine learning community are inductive models, one of the most discussed useful biases is *inductive bias* [32]. For instance, PK on protein interactions can impact inference on activation cascades; only upstream proteins can activate downstream proteins, not vice versa.

## Why do we need biases?

Humans will be the gold standard for common-sense reasoning for the foreseeable future. However, human reasoning is limited by our sensory and mnemonic capacity; we cannot reason about high-dimensional data since we can neither perceive it nor keep it in memory. Machine learning seems like the ideal solution, but the "No Free Lunch" theorems present a fundamental challenge: no single learning algorithm may be universally superior across all problem domains [33]. Although they have recently been challenged [34], these theorems highlight the inherent difficulty in designing algorithms that generalise well from specific training data to new, unseen data. Inductive biases guide algorithms in making educated guesses about unseen data, thereby improving their generalisation capabilities [35].

This need for inductive biases is particularly apparent in the realm of biomedicine [36]. Biomedical research operates within a framework constrained by limited and often high-dimensional data, stemming from the high costs of experiments, the scarcity of samples, and the inherent complexity of biological systems. Coupled with the natural variability of biological measurements, these factors result in a low signal-to-noise ratio, making it challenging to discern meaningful patterns. Inductive biases direct the learning process towards more relevant solutions by incorporating assumptions that enable more effective learning and interpretation, ensuring that models are not just statistically sound but also biologically meaningful.

Some central questions then arise:

- How explicit should we be in introducing biases, i.e., should the model determine its own biases, or do we force them on the model?

- How do we choose the right biases to introduce?

- How do we evaluate the biases we introduce?

## Bias from prior knowledge

The first question alone is highly debated in the wider field of machine learning; it is related to the concept of the bias-variance tradeoff. The frequently quoted "Bitter Lesson" posits that we should refrain from inducing all but the most basic biases in our models, and that we should not view metrics as the ultimate measure of performance, but rather whether the model gets us closer to some truth [37]. However, it has been argued that many improvements that led to the models of today, such as convolution or attention, disprove this theory [38], and that the intrinsic complexity of real-world systems does not obviate, but rather necessitate, the integration of human insight into our learning frameworks [39,40].

In systems biology, specifically, there is much interest in finding models with suitable biases to deal with constraints specific to the field, such as data availability and the completeness of PK

[9,35,41,42,43]. Considering these constraints, the question is not whether to include PK in our reasoning, but which knowledge, when, and how [40].

## Prior knowledge

PK refers to information or data that is available to inform a learning process, enhancing the performance of the trained models and their ability to generalise. It can be used to inform the inductive biases of a model, either explicitly through the design choices and assumptions embedded into the models, or implicitly through the data and methods used in training. For this to be possible, biomedical entities and relationships must be clearly defined and represented unambiguously. Additionally, the diversity in our tasks and knowledge sources requires a flexible representation. Knowledge representation frameworks can aid in this process [44].

In the biomedical field, there is a rich tradition of documenting biological knowledge at various levels of detail and focusing on different aspects of biology. Detailed mechanistic models provide mathematical descriptions of the dynamic interactions at a molecular or cellular level. Genome-scale networks, including metabolic and gene regulatory networks, offer comprehensive views of metabolic processes and gene interactions [45]. Protein-protein interaction databases recapitulate either causal or non-causal interactions between proteins [45].

## Modelling on prior knowledge

The integration of PK into models is a non-trivial but essential process for moving from correlation to causation. PK can be used to derive inductive biases either *explicitly* or *implicitly*.

The explicit case typically involves a mathematical framework where a set of assumptions is explicitly stated and integrated into the model. Ordinary Differential Equation (ODE) models, logic-based models, rule-based models, and constraint-based models [46], all of which are commonly used in systems biology, explicitly incorporate different types of PK, can be fitted to data, and then be used to answer different types of causal questions. In the field of CR, Structural Causal Models can be used when mechanisms are unknown [6,12]. Their advantage is high efficiency in the face of scarce data, but they are highly reliant on the quality and comprehensiveness of the underlying PK [47].

In contrast, implicit integration of PK in models involves learning useful representations directly from the data, without the explicit inclusion of biological assumptions or causal knowledge. Learning mechanisms introduced as implicit biases can be simple (e.g., sparsity) or elaborate. Simple implicit biases include regularisation techniques that help models generalise by preventing overfitting [48], or decisions about the types of prior distributions in bayesian models [49]. More elaborate are neural networks which employ specific architectural designs, such as Convolutional Neural Networks (CNNs) [50], Recurrent Neural Networks (RNNs) [51], or Transformers [38]. Their advantages and disadvantages are inverse to those of explicit models [47].

As a result, choosing the best way to derive inductive biases from PK is not straightforward. Models that explicitly incorporate PK are more interpretable and can generalise effectively even when data is scarce [47]. However, they are constrained by the accuracy of the existing knowledge and often struggle to scale to larger datasets [52,53]. Models with implicit biases, on the other hand, particularly those typically found in deep learning architectures, excel at learning from large, high-dimensional datasets and offer flexibility across diverse domains. Yet, they suffer from limited interpretability, are prone to overfitting, and typically do not generalise well to scenarios not encountered during training, such as predicting the effects of new drugs or drug combinations, largely due to their lack of causal knowledge.

Hybrid models make a tradeoff between those extremes, which is why they have been found to be useful in systems biology, where data are currently scarce [54,55,56,57,58,59,60]. While their implementation details differ, they often employ two learners side-by-side, one of which is driven by explicit biases from PK, while the other learns from data. Frequently, these learners are also coupled in an end-to-end learning process, i.e., they "learn together." This mode of learning aims to benefit from the "bias-free" nature of neural networks while simultaneously improving model performance in the face of scarce data via the added explicit bias.

# Causality in foundation models

There has been an enormous spike of interest in attention-based neural network models, in large part due to the success of Large Language Models (LLMs). While the high performance of LLMs is based on myriad technical improvements, the introduction of attention as an architectural bias has been a major contributor to their success [38]. This has led to the development of attention-based molecular models (most commonly for gene expression), which can also be considered "GPT" models: Generative Pre-trained Transformers [61,62,63]. Attention as a learning mechanism enables the integration of non-local information in a flexible manner. In a molecular model that reasons about gene expression, such as Geneformer, attention allows the integration of distant regulatory elements [62]. Notably, this mechanism comes with a computational cost that increases exponentially with respect to the length of the input sequence [64].

The generalist capabilities of LLMs have led to the designation of "foundation models" [65]. Foundation models are models that achieve high performance by training a generic architecture on extremely large amounts of data in a self-supervised manner. They can be fine-tuned for more specific tasks, because they are thought to derive generalisable representations and mechanisms by training on an amount of data large enough to learn the complexity of real-world systems. However, recent molecular foundation model benchmarks highlight clear discrepancies between the "foundational" aspirations of the pre-trained models and the real-world evaluation of their performance [66,67]. Briefly, the benchmarks found that, on single cell classification tasks, the proposed foundation models did not outperform simple baselines consistently when applied "zero-shot," i.e., without fine-tuning. State-of-the-art methods such as scVI [68] and even the mere selection of highly variable genes was often statistically indistinguishable from the highly parameterised methods, and sometimes even yielded better classification outcomes. However, these are early models, and it could still be argued that, in line with the scaling hypothesis, models may improve via a combination of the right architecture with sufficient amounts of data [69].

Indeed, molecular foundation models lag behind in size: while current-generation LLMs have around 100 billion parameters or more and are trained on enormous text corpuses (hundreds of billions to trillions of tokens), molecular foundation models have tens of millions of parameters (scGPT: 53M, Geneformer: 10M) and are trained on corpuses of tens of millions of cells, which (optimistically) yields hundreds of billions of individual data points. Thus, LLMs are currently about 2000 times larger than molecular foundation models, while arguably also dealing with a less complicated system. The question whether scaling will lead to the emergence of "foundational behaviour" in molecular models is still a matter of much debate [70].

## Attention - and large amounts of data - is all you need?

Given enough data to train on - and ample funds for compute - is attention "all you need" to induce reliable biases in your model? While there are doubts regarding the reasoning capabilities of LLMs, GPT arguably "understands" language very well already, to the point where it can flawlessly communicate and synthesise information [71]. This is what the term "foundation model" implies: the model has derived a generalisable representation of language, a tool that can be fine-tuned for a variety of language-related tasks. This behaviour is not possible without assuming some form of causality, even if it is not explicitly encoded in the model [17].

In this light, what are the reasons to be sceptical about the capacity of molecular foundation models to understand the "grammar" of the cell?

**Explainability**: For one, large transformer models (i.e., billions of parameters) are not explainable due to their high complexity. As such, there is often no way to scrutinise their reasoning beyond the output they produce [72,73]. What seems simple in the case of language models - the famous Turing test can be performed by any human with a basic understanding of language - is exceedingly difficult in the molecular space, where many causal relationships are still unknown [71]. Yet the only way to scrutinise and subsequently improve the reasoning capabilities of a model is precisely this explicit validation of its predictions in an interpretable setting.

While the creation of explicit molecular models (e.g., logic, structural causal, or ODE-based models) and the self-supervised training of molecular foundation models are methodically very different, both can provide a hypothesis on causal structure that can be formulated as a network. Theodoris et al. explore the attention layers of their Geneformer foundation model to explain the model's reasoning [62]. While some layers show clear patterns of attention, such as attending to highly connected or highly expressed genes, other layers are not as readily interpretable, much less so than explicit molecular models.

**Benchmarking**: Whether these complex layers reflect the true complexity of the underlying biology or are rather evidence for overfitting to the training data is not clear. One argument in favour of overfitting is the poor generalisation of the model in independent benchmarks [66,67]. To determine whether molecular foundation models indeed capture generalisable causal representations of biology, dedicated benchmarks are needed.

**Causal bias**: The GPT-3 architecture that led to the recent breakthrough in LLM capabilities employs "causal self-attention," describing an implicit architectural bias that prevents the model from "looking into the future": for predicting the next token, only the previous tokens in the sentence can be used [64]. This leverages the implicit causality present in language, which incidentally is similar to one of the earliest formal descriptions of causality, that "the effect has regularly followed the cause in the past" [74]. Compared to language, the data that form the input of molecular foundation models do not implicitly contain causal information. The individual cells are in general not on a known trajectory, and the genes that are masked as part of the training objective are masked at random, not because they are downstream (in some form) of the genes used for prediction. This fundamental difference between language and molecular models has so far not been explored theoretically or empirically.

## Causal latent spaces

Due to the fundamental limitation of human perception, dimensionality reduction is a popular workflow for data interpretation, typically via methods such as PCA, t-SNE, or UMAP [75]. The hope is that exploration and explanation in the lower-dimensional embedding space may be less challenging than in the original data, which assumes that the most important aspects of variability in the original data are captured in the reduced dimensions [76]. However, without explicit supervision, which is rare in typical biomedical datasets, the resulting latent spaces are rarely interpretable, and do not lend themselves to causal interpretation. In addition, they often suffer from biases that result from technical rather than biological factors [77]. In consequence, biological insight during the exploration of these latent spaces is often challenging due to the dominance of biases over the biological generative mechanism.

Performing causal inference in latent spaces could potentially solve some of these issues. "Moving through the latent space" reduces the number of variables that change upon intervention, making exploration simpler in theory. In practice, however, ease and sensibility of exploration depend completely on whether the inductive biases in the embedding process capture the underlying biology. In addition, latent spaces have no trivial connection to the real-world measurements they are based on. Each model instance generates its own, independent latent space; in consequence, the exploration of latent spaces is challenging and time-consuming.

Even if a given latent space can be explored, there is often no guarantee that interpolation between sensible latent representations also leads to sensible results. As an example, consider a prevailing issue of visual generative models in drawing human hands: images of hands typically involve mangled anatomy and an incorrect number of digits [78]. Even though there is a section in the latent space that represents hands, this does not represent the concept of a hand, but rather is guided by learning on many diverse pictures of hands. A section of this latent space may represent only a finger, and carry some information that next to a finger there usually is another finger. However, when generating the image, there is no mechanism to keep track of how many digits to add to any generated hand, leading to wrong anatomy. Similarly, when exploring the latent space of a model of molecular signalling, there may be no guarantees that the model respects the concept of a given pathway when generating the signalling molecules involved.

If mastered, exploring and performing interventions in latent spaces promises many benefits: better generalisation and improved sample efficiency [42], predicting the outcomes of interventions not observed at training time [79], or insights into the effect of different inductive biases in the model [80]. However, to achieve this, gaining a better understanding of properties of the learned embeddings and variables is essential, for instance by performing "imagined interventions" in the latent space [81] or by using model uncertainty for guiding the optimisation process in the latent space [82]. Of note, many of the proposed solutions for more explainable latent spaces depend on architectures that may scale significantly worse than transformers [52,53].

# Discussion / Conclusion

## Dichotomy of scaling (data-driven) and bias injection (knowledge-driven)

The debate between adopting scaling strategies versus the injection of biases from prior knowledge highlights a fundamental tension in modern biomedical research. The "Bitter Lesson" suggests a preference for general-purpose learning algorithms that scale with computational resources, implicitly learning biases from data. However, complex models often pose significant computational challenges; many models are limited to network sizes unfeasibly small for biological inference, and feedback loops are often excluded [12]. Conversely, explicitly injecting biases from PK can lead to more specialised and efficient models that can generalise using relatively little training data, but may not scale. Hybrid models represent a promising middle ground, combining the scalability of generalist models with the efficiency and specificity provided by tailored biases. Researchers often rely on intuition to determine which biases to inject, understanding that while no single model may universally excel (reflecting the "No Free Lunch" theorems), the blend of generalisation through scaling and specialisation through bias injection might provide a robust framework for tackling complex biomedical challenges.

## Theoretical foundations: interventions and inductive biases

Theoretical work emphasises the need for interventions in causal discovery but does not yet address the influence of inductive biases [83]. The number of required interventions might be reduced significantly when complemented with high-quality observational data and appropriate biases, as suggested by neural causal models [84]. However, the precise nature of these biases and their impact remains understudied theoretically as well as empirically. The comparative effectiveness and theoretical underpinnings of explicit models versus implicit models are particularly understudied. Foundation models have embraced causal self-attention as a step towards integrating causality, but this alone may be insufficient. Empirical studies and more robust theories are needed to understand these dynamics, including the potential existence and validation of causal latent spaces.

## Data types: observational vs. interventional

The choice between training on observational versus interventional data (or a mixture of both) is critical in the development of models. While large-scale data collection is vital, the type of data collected can significantly influence model performance and the ability to generalise and make accurate causal inferences. The complexity and high cost of collecting data requires an efficient experimental design to maximise causal discovery with limited resources. Observational data are more readily available but may lead to confounded or biassed insights. Interventional data, while more challenging to obtain, provide clearer causal pathways and can greatly enhance the model's understanding of underlying biological processes [85,86]. A balanced approach, possibly incorporating both observational and interventional data, coupled with mechanisms for deciding the right number and type of interventions, might provide a more nuanced understanding and improve model robustness and interpretability. In addition, experimental design decisions such as the inclusion of a temporal axis can improve the amenability of observational data to causal inference.

## Foundation models: architectural biases and the No Free Lunch theorems

Foundation models challenge the "No Free Lunch" theorems by suggesting that certain architectural biases, learned from vast amounts of data, can yield generalisable and high-performing models [34]. These biases, and how to transfer them from LLMs to systems biology, necessitate careful evaluation. As the biomedical field looks to these models for answers, it becomes crucial to develop frameworks that facilitate rapid development and exploration of ideas [44,87]. A crucial aspect of these frameworks will be establishing benchmarks in the face of missing biological ground truth.

## Systems biology and causality - finding a balance

Systems biology has historically followed both knowledge-driven (bottom-up) and data-driven (top-down) approaches. Bottom-up systems biology, aiming to understand specific molecular mechanisms driving biological phenomena, has *de facto* been doing CR, despite both fields being largely disconnected. Meanwhile, top-down systems biology, inspired more by machine learning principles, has struggled with moving from correlation to causality. New methods and models offer the potential to converge these complementary approaches and scale our understanding to larger, more complex systems. However, it remains to be seen whether the future of biological modelling will be dominated by the generation of vast datasets for generalist models or by more nuanced, bias-inclusive architectures.

While the allure of generalist models trained on extensive datasets is strong, the unique challenges of biomedical research may necessitate a more tailored approach. Including explicit favourable biases, informed by deep domain knowledge and specific data types (observational or interventional), could lead to breakthroughs in understanding complex biological systems. The field must explore these possibilities, balancing the drive for large-scale data with the need for precision and specificity, to realise the full potential of modern systems biology.

# Acknowledgements

We thank Aurelien Dugourd, Philipp Schäfer, Loan Vulliard, and Jan Lanzer for their helpful comments on the manuscript.

# Funding

This work was supported by the European Union's Horizon 2020 Programme under PerMedCoE (951773) and DECIDER (965193).

# Conflict of Interest

JSR reports funding from GSK, Pfizer, and Sanofi and fees/honoraria from Travere Therapeutics, Stadapharm, Astex, Pfizer, and Grunenthal.

# Glossary

## Attention (deep learning)

A mechanism in deep learning that allows the model to focus on specific parts of the input data. Attention mechanisms are often used in natural language processing to focus on specific words in a sentence, but can also be used in other domains.

## Bias (machine learning)

Bias can be understood in two ways in the context of machine learning.

1. The first definition, and the one predominantly used in this Perspective, is also referred to as statistical bias; a technical term referring to the assumptions made by a model to make predictions. This bias is a necessary part of any machine learning model. A model with high bias (low variance) pays very little attention to the training data and oversimplifies the model, which can lead to [underfitting](). This means it does not capture the complexity of the data and fails to learn the underlying patterns effectively. Conversely, a model with low bias (high variance) makes complex assumptions to fit the data closely, which can lead to [overfitting](), where the model captures noise in the data as if it were a true pattern. See also the [bias-variance tradeoff]().

2. The second definition is also known as algorithmic bias, and refers to the systematic and repeatable errors in a model due to faulty assumptions or data. It often reflects existing biases in the real world that the training data are derived from, but can also result from architectural choices in the model. As such, algorithmic bias can result from any stage in model training, from data collection to model deployment.

## Bias-variance tradeoff

The concept in machine learning that [bias]() and [variance]() of a model are inversely related. The term implies that an optimal model finds a balance between bias (impact of the model on predictions) and variance (impact of the data on predictions). This balance depends on the complexity of model and data.

## Deductive vs. Inductive Reasoning

Deductive reasoning involves drawing specific conclusions from general statements or premises, whereas inductive reasoning involves making broad generalisations from specific observations. Deductive reasoning is often seen as more logically sound but less informative about the real world, while inductive reasoning is more exploratory but can lead to less certain conclusions.

## Do-Calculus

Developed by Judea Pearl, Do-Calculus is a formal mathematical framework used in causal inference. It provides a set of rules for calculating the effects of interventions in probabilistic models, allowing researchers to infer causality from observational data.

## Foundation model

A model that is trained on a large amount of data and can be used as a starting point for further model development (also referred to as fine-tuning). Foundation models are assumed to have learned generalisable patterns from their input data. To achieve this, they require large amounts of data and compute power.

## Large Language Models

Large Language Models are advanced AI models trained on extensive text data. They are capable of understanding and generating human-like text, making them useful in various applications like translation, summarization, and conversation. LLMs leverage vast amounts of training data to grasp nuances of language, context, and even some elements of human communication. They are the first commercially successful example of foundation models.

## 'No Free Lunch' Theorems

These theorems in optimization and machine learning suggest that no single algorithm is best for every problem. The performance of an algorithm is contingent on the specificities of the task and data at hand. This highlights the importance of choosing or designing algorithms that are well-suited to the particular characteristics of the problem being addressed. Related to the [bias-variance tradeoff](), partly opposed to the [scaling hypothesis]() and [foundation models]().

## Overfitting

A technical term referring to a model that captures noise in the data as if it were a true pattern. Overfitting tends to lead to high performance on the training data but poor performance on the test data. If a model has overfitted also to the test data, it will also perform poorly on new data, i.e., it will not generalise well. This is a common occurrence if there has been data leakage between training, validation, and test data.

## Prior knowledge

A term referring to information that is available to inform a learning process. Often, this is the result of previous research.

## Randomised Clinical Trials

Randomised clinical trials are experiments designed to test the efficacy of medical interventions. Participants are randomly assigned to groups receiving different treatments, including a control group, which often receives a placebo or gold-standard treatment. To further minimise confounding factors, participants and administering doctors are often blinded to the treatment given. This method is considered the gold standard in clinical research for its ability to minimise bias and establish causality between a treatment and its outcomes.

## Scaling hypothesis

The scaling hypothesis posits that the performance of a model increases with the amount of data it is trained on. Recently, it has come to describe the idea that, given enough data, complex model behaviours can emerge. The enormous success of current Large Language Models has been attributed to scaling, with emergence of human-like language capabilities around the time of GPT-3. The ability to scale depends on several factors: the availability of data, parallelisation of training, adequate compute power with a parallel architecture, and a model architecture that can digest large amounts of data effectively.

## Self-supervised learning

A type of machine learning where the model learns from the data itself, without the need for human labelling. This is achieved by training the model to predict certain properties of the data, such as the next word in a sentence, or the next frame in a video. Self-supervised learning is often used in the pre-training of [foundation models](). It typically requires a specific mechanism in the model to account for the lack of labelled data, such as the masking applied in the training of [Large Language Models]().

## Structural Causal Models (SCMs)

SCMs are a type of statistical model used to represent and analyse causal relationships. They consist of variables and equations that describe how these variables interact causally. SCMs are particularly useful in causal inference as they allow for the analysis of how changes in one variable may cause changes in another.

## Underfitting

A technical term referring to a model that does not capture the complexity of the data. Underfitting tends to lead to poor performance on both the training and test data.

## Variance (machine learning)

A technical term referring to the sensitivity of a model to the training data. Describes how much the predictions of a model vary given different training data. High variance (low bias) in a model can lead to [overfitting]() and thus harm generalisation. Conversely, low variance (high bias) can lead to [underfitting]() and thus to a model that does not capture the complexity of the data. See also the [bias-variance tradeoff]().

# References

1.  **ORGANON OR LOGICAL TREATISES O**
    Aristotle, Octavius Freire 1816?-1873 Owen
    *Wentworth Press* (2016)
    ISBN: 9781372537233

2.  **Causality**
    Judea Pearl
    *Cambridge University Press* (2009-09-14) https://doi.org/ggd72q
    DOI: 10.1017/cbo9780511803161

3.  **Identification of Causal Effects Using Instrumental Variables**
    Joshua D Angrist, Guido W Imbens, Donald B Rubin
    *Journal of the American Statistical Association* (1996-06) https://doi.org/gdz4f4
    DOI: 10.1080/01621459.1996.10476902

4.  **Myth and measurement: the new economics of the minimum wage**
    David E Card, Alan B Krueger
    *Princeton University Press* (2016)
    ISBN: 9781400880874

5.  **Causal Machine Learning: A Survey and Open Problems**
    Jean Kaddour, Aengus Lynch, Qi Liu, Matt J Kusner, Ricardo Silva
    (2022) https://arxiv.org/abs/2206.15475

6.  **Causal machine learning for single-cell genomics**
    Alejandro Tejada-Lapuerta, Paul Bertin, Stefan Bauer, Hananeh Aliee, Yoshua Bengio, Fabian J Theis
    *arXiv* (2023) https://doi.org/gtb97p
    DOI: 10.48550/arxiv.2310.14935

7.  **A map of human genome variation from population-scale sequencing** *Nature* (2010-10-27)
    https://doi.org/fmk7rw
    DOI: 10.1038/nature09534 · PMID: 20981092 · PMCID: PMC3042601

8.  **Causality in digital medicine**
    Nature Communications
    (2021-09-15) https://doi.org/gtcfvh
    DOI: 10.1038/s41467-021-25743-9 · PMID: 34526509 · PMCID: PMC8443583

9.  **The perpetual motion machine of AI-generated data and the distraction of ChatGPT-as-scientist**
    Jennifer Listgarten
    *arXiv* (2023) https://doi.org/gs8pnp
    DOI: 10.48550/arxiv.2312.00818

10. **The Do-Calculus Revisited**
    Judea Pearl
    *arXiv* (2012) https://doi.org/gtbf4r
    DOI: 10.48550/arxiv.1210.4852

11. **Improved Survival with Vemurafenib in Melanoma with BRAF V600E Mutation**

Paul B Chapman, Axel Hauschild, Caroline Robert, John B Haanen, Paolo Ascierto, James Larkin, Reinhard Dummer, Claus Garbe, Alessandro Testori, Michele Maio, … Grant A McArthur
*New England Journal of Medicine* (2011-06-30) https://doi.org/dsbxxt
DOI: 10.1056/nejmoa1103782 · PMID: 21639808 · PMCID: PMC3549296

12. **Causal Structure Learning: A Combinatorial Perspective**
    Chandler Squires, Caroline Uhler
    *Foundations of Computational Mathematics* (2022-08-01) https://doi.org/gtdtds
    DOI: 10.1007/s10208-022-09581-9 · PMID: 35935470 · PMCID: PMC9342837

13. **Reliable interpretability of biology-inspired deep neural networks**
    Wolfgang Esser-Skala, Nikolaus Fortelny
    *npj Systems Biology and Applications* (2023-10-10) https://doi.org/gtb95c
    DOI: 10.1038/s41540-023-00310-8 · PMID: 37816807 · PMCID: PMC10564878

14. **Structural Identifiability of Systems Biology Models: A Critical Comparison of Methods**
    Oana-Teodora Chis, Julio R Banga, Eva Balsa-Canto
    *PLoS ONE* (2011-11-22) https://doi.org/fch6rc
    DOI: 10.1371/journal.pone.0027755 · PMID: 22132135 · PMCID: PMC3222653

15. **The role of causality in explainable artificial intelligence**
    Gianluca Carloni, Andrea Berti, Sara Colantonio
    *arXiv* (2023) https://doi.org/gtb95k
    DOI: 10.48550/arxiv.2309.09901

16. **Causal Structure Learning**
    Christina Heinze-Deml, Marloes H Maathuis, Nicolai Meinshausen
    *Annual Review of Statistics and Its Application* (2018-03-07) https://doi.org/ggh4pj
    DOI: 10.1146/annurev-statistics-031017-100630

17. **Can Foundation Models Talk Causality?**
    Moritz Willig, Matej Zečević, Devendra Singh Dhami, Kristian Kersting
    *arXiv* (2022) https://doi.org/gtb9wb
    DOI: 10.48550/arxiv.2206.10591

18. **The Scaling Hypothesis**
    Gwern Branwen
    (2020-05-28) https://gwern.net/scaling-hypothesis

19. **Inferring causal molecular networks: empirical assessment through a community-based effort**
    Steven M Hill, Laura M Heiser, Thomas Cokelaer, Michael Unger, Nicole K Nesser, Daniel E Carlin, Yang Zhang, Artem Sokolov, Evan O Paull, … Sach Mukherjee
    *Nature Methods* (2016-02-22) https://doi.org/f3t7t4
    DOI: 10.1038/nmeth.3773 · PMID: 26901648 · PMCID: PMC4854847

20. **Integrating knowledge and omics to decipher mechanisms via large-scale models of signaling networks**
    Martin Garrido-Rodriguez, Katharina Zirngibl, Olga Ivanova, Sebastian Lobentanzer, Julio Saez-Rodriguez
    *Molecular Systems Biology* (2022-07) https://doi.org/gtb9v8
    DOI: 10.15252/msb.202211036 · PMID: 35880747 · PMCID: PMC9316933

21. **Illuminating the dark phosphoproteome**
    Elise J Needham, Benjamin L Parker, Timur Burykin, David E James, Sean J Humphrey
    *Science Signaling* (2019-01-22) https://doi.org/gf8c3h

DOI: 10.1126/scisignal.aau8645 · PMID: 30670635

22. **The functional landscape of the human phosphoproteome**
    David Ochoa, Andrew F Jarnuczak, Cristina Viéitez, Maja Gehre, Margaret Soucheray, André Mateus, Askar A Kleefeldt, Anthony Hill, Luz Garcia-Alonso, Frank Stein, … Pedro Beltrao
    *Nature Biotechnology* (2019-12-09) https://doi.org/ggd8n7
    DOI: 10.1038/s41587-019-0344-3 · PMID: 31819260 · PMCID: PMC7100915

23. **The book of why: the new science of cause and effect**
    Judea Pearl, Dana Mackenzie
    *Basic Books* (2018)
    ISBN: 9780465097616

24. **Causal inference in statistics: An overview**
    Judea Pearl
    *Statistics Surveys* (2009-01-01) https://doi.org/drt748
    DOI: 10.1214/09-ss057

25. **Regression discontinuity designs: A guide to practice**
    Guido W Imbens, Thomas Lemieux
    *Journal of Econometrics* (2008-02) https://doi.org/bzx7rb
    DOI: 10.1016/j.jeconom.2007.05.001

26. **Perturb-Seq: Dissecting Molecular Circuits with Scalable Single-Cell RNA Profiling of Pooled Genetic Screens**
    Atray Dixit, Oren Parnas, Biyu Li, Jenny Chen, Charles P Fulco, Livnat Jerby-Arnon, Nemanja D Marjanovic, Danielle Dionne, Tyler Burks, Raktima Raychowdhury, … Aviv Regev
    *Cell* (2016-12) https://doi.org/f9prjd
    DOI: 10.1016/j.cell.2016.11.038 · PMID: 27984732 · PMCID: PMC5181115

27. **Vemurafenib: the first drug approved for BRAF-mutant cancer**
    Gideon Bollag, James Tsai, Jiazhong Zhang, Chao Zhang, Prabha Ibrahim, Keith Nolop, Peter Hirth
    *Nature Reviews Drug Discovery* (2012-10-12) https://doi.org/f4b975
    DOI: 10.1038/nrd3847 · PMID: 23060265

28. **Targeting the ERK Signaling Pathway in Melanoma**
    Paola Savoia, Paolo Fava, Filippo Casoni, Ottavio Cremona
    *International Journal of Molecular Sciences* (2019-03-25) https://doi.org/gtb9v9
    DOI: 10.3390/ijms20061483 · PMID: 30934534 · PMCID: PMC6472057

29. **A Theory of Causal Learning in Children: Causal Maps and Bayes Nets.**
    Alison Gopnik, Clark Glymour, David M Sobel, Laura E Schulz, Tamar Kushnir, David Danks
    *Psychological Review* (2004) https://doi.org/fkj59s
    DOI: 10.1037/0033-295x.111.1.3 · PMID: 14756583

30. **How to Grow a Mind: Statistics, Structure, and Abstraction**
    Joshua B Tenenbaum, Charles Kemp, Thomas L Griffiths, Noah D Goodman
    *Science* (2011-03-11) https://doi.org/d8vvm9
    DOI: 10.1126/science.1192788 · PMID: 21393536

31. **A Survey on Bias and Fairness in Machine Learning**
    Ninareh Mehrabi, Fred Morstatter, Nripsuta Saxena, Kristina Lerman, Aram Galstyan
    *arXiv* (2019) https://doi.org/gtb9wn
    DOI: 10.48550/arxiv.1908.09635

32. **A Model of Inductive Bias Learning**
    J Baxter
    *Journal of Artificial Intelligence Research* (2000-03-01) https://doi.org/gg66h8
    DOI: 10.1613/jair.731

33. **No Free Lunch Theorems for Search**
    David H Wolpert, William G Macready
    *Santa Fe Institute* (1995) https://ideas.repec.org/p/wop/safiwp/95-02-010.html

34. **The No Free Lunch Theorem, Kolmogorov Complexity, and the Role of Inductive Biases in Machine Learning**
    Micah Goldblum, Marc Finzi, Keefer Rowan, Andrew Gordon Wilson
    *arXiv* (2023) https://doi.org/gtb9wp
    DOI: 10.48550/arxiv.2304.05366

35. **Inductive biases for deep learning of higher-level cognition**
    Anirudh Goyal, Yoshua Bengio
    *Proceedings of the Royal Society A: Mathematical, Physical and Engineering Sciences* (2022-10) https://doi.org/gs39f8
    DOI: 10.1098/rspa.2021.0068

36. **Current progress and open challenges for applying deep learning across the biosciences**
    Nicolae Sapoval, Amirali Aghazadeh, Michael G Nute, Dinler A Antunes, Advait Balaji, Richard Baraniuk, CJ Barberan, Ruth Dannenfelser, Chen Dun, Mohammadamin Edrisi, … Todd J Treangen
    *Nature Communications* (2022-04-01) https://doi.org/gp26xk
    DOI: 10.1038/s41467-022-29268-7 · PMID: 35365602 · PMCID: PMC8976012

37. **The Bitter Lesson** http://www.incompleteideas.net/IncIdeas/BitterLesson.html

38. **Attention Is All You Need**
    Ashish Vaswani, Noam Shazeer, Niki Parmar, Jakob Uszkoreit, Llion Jones, Aidan N Gomez, Lukasz Kaiser, Illia Polosukhin
    *arXiv* (2017) https://doi.org/gpnmtv
    DOI: 10.48550/arxiv.1706.03762

39. **A Better Lesson – Rodney Brooks** (2019-03-19) https://rodneybrooks.com/a-better-lesson/

40. **Thread by @shimon8282: "Rich Sutton has a new blog post entitled "The Bitter Lesson" (incompleteideas.net/IncIdeas/Bitte…) that I strongly disagree with. In it, he […]"**
    https://twitter.com/shimon8282
    https://threadreaderapp.com/thread/1106534178676506624.html

41. **Challenging Common Assumptions in the Unsupervised Learning of Disentangled Representations**
    Francesco Locatello, Stefan Bauer, Mario Lucic, Gunnar Rätsch, Sylvain Gelly, Bernhard Schölkopf, Olivier Bachem
    *arXiv* (2018) https://doi.org/grx79c
    DOI: 10.48550/arxiv.1811.12359

42. **Toward Causal Representation Learning**
    Bernhard Scholkopf, Francesco Locatello, Stefan Bauer, Nan Rosemary Ke, Nal Kalchbrenner, Anirudh Goyal, Yoshua Bengio
    *Proceedings of the IEEE* (2021-05) https://doi.org/gjhqrh
    DOI: 10.1109/jproc.2021.3058954

43. **Beyond Predictions in Neural ODEs: Identification and Interventions**
    Hananeh Aliee, Fabian J Theis, Niki Kilbertus
    *arXiv* (2021) https://doi.org/gszw4d
    DOI: 10.48550/arxiv.2106.12430

44. **Democratizing knowledge representation with BioCypher**
    Sebastian Lobentanzer, Patrick Aloy, Jan Baumbach, Balazs Bohar, Vincent J Carey, Pornpimol Charoentong, Katharina Danhauser, Tunca Doğan, Johann Dreo, Ian Dunham, … Julio Saez-Rodriguez
    *Nature Biotechnology* (2023-06-19) https://doi.org/gszqjr
    DOI: 10.1038/s41587-023-01848-y · PMID: 37337100

45. **Quantitative and logic modelling of molecular and gene networks**
    Nicolas Le Novère
    *Nature Reviews Genetics* (2015-02-03) https://doi.org/f6299z
    DOI: 10.1038/nrg3885 · PMID: 25645874 · PMCID: PMC4604653

46. **Constraint-based models predict metabolic and associated cellular functions**
    Aarash Bordbar, Jonathan M Monk, Zachary A King, Bernhard O Palsson
    *Nature Reviews Genetics* (2014-01-16) https://doi.org/f5sk8s
    DOI: 10.1038/nrg3643 · PMID: 24430943

47. **Model scale versus domain knowledge in statistical forecasting of chaotic systems**
    William Gilpin
    *Physical Review Research* (2023-12-15) https://doi.org/gs9xz3
    DOI: 10.1103/physrevresearch.5.043252

48. **Regression Shrinkage and Selection Via the Lasso**
    Robert Tibshirani
    *Journal of the Royal Statistical Society: Series B (Methodological)* (1996-01)
    https://doi.org/gfn45m
    DOI: 10.1111/j.2517-6161.1996.tb02080.x

49. **A general and flexible method for signal extraction from single-cell RNA-seq data**
    Davide Risso, Fanny Perraudeau, Svetlana Gribkova, Sandrine Dudoit, Jean-Philippe Vert
    *Nature Communications* (2018-01-18) https://doi.org/gcv5w7
    DOI: 10.1038/s41467-017-02554-5 · PMID: 29348443 · PMCID: PMC5773593

50. **Backpropagation Applied to Handwritten Zip Code Recognition**
    Y LeCun, B Boser, JS Denker, D Henderson, RE Howard, W Hubbard, LD Jackel
    *Neural Computation* (1989-12) https://doi.org/bknd8g
    DOI: 10.1162/neco.1989.1.4.541

51. **Long Short-Term Memory**
    Sepp Hochreiter, Jürgen Schmidhuber
    *Neural Computation* (1997-11-01) https://doi.org/bxd65w
    DOI: 10.1162/neco.1997.9.8.1735 · PMID: 9377276

52. **Scaling Laws for Neural Language Models**
    Jared Kaplan, Sam McCandlish, Tom Henighan, Tom B Brown, Benjamin Chess, Rewon Child, Scott Gray, Alec Radford, Jeffrey Wu, Dario Amodei
    *arXiv* (2020) https://doi.org/gtb96w
    DOI: 10.48550/arxiv.2001.08361

53. **Investigating Power laws in Deep Representation Learning**
    Arna Ghosh, Arnab Kumar Mondal, Kumar Krishna Agrawal, Blake Richards

*arXiv* (2022) https://doi.org/gtb966
DOI: 10.48550/arxiv.2202.05808

54. **Differentiable biology: using deep learning for biophysics-based and data-driven modeling of molecular mechanisms**
Mohammed AlQuraishi, Peter K Sorger
*Nature Methods* (2021-10) https://doi.org/gm2b58
DOI: 10.1038/s41592-021-01283-4 · PMID: 34608321 · PMCID: PMC8793939

55. **Artificial neural networks enable genome-scale simulations of intracellular signaling**
Avlant Nilsson, Joshua M Peters, Nikolaos Meimetis, Bryan Bryson, Douglas A Lauffenburger
*Nature Communications* (2022-06-02) https://doi.org/gqd9j9
DOI: 10.1038/s41467-022-30684-y · PMID: 35654811 · PMCID: PMC9163072

56. **A neural-mechanistic hybrid approach improving the predictive power of genome-scale metabolic models**
Léon Faure, Bastien Mollet, Wolfram Liebermeister, Jean-Loup Faulon
*Nature Communications* (2023-08-03) https://doi.org/gtb96p
DOI: 10.1038/s41467-023-40380-0 · PMID: 37537192 · PMCID: PMC10400647

57. **Predicting transcriptional outcomes of novel multigene perturbations with GEARS**
Yusuf Roohani, Kexin Huang, Jure Leskovec
*Nature Biotechnology* (2023-08-17) https://doi.org/gtb96r
DOI: 10.1038/s41587-023-01905-6 · PMID: 37592036

58. **Knowledge-primed neural networks enable biologically interpretable deep learning on single-cell sequencing data**
Nikolaus Fortelny, Christoph Bock
*Genome Biology* (2020-08-03) https://doi.org/gg8ws9
DOI: 10.1186/s13059-020-02100-5 · PMID: 32746932 · PMCID: PMC7397672

59. **Biologically informed deep learning to query gene programs in single-cell atlases**
Mohammad Lotfollahi, Sergei Rybakov, Karin Hrovatin, Soroor Hediyeh-zadeh, Carlos Talavera-López, Alexander V Misharin, Fabian J Theis
*Nature Cell Biology* (2023-02-02) https://doi.org/gtb96q
DOI: 10.1038/s41556-022-01072-x · PMID: 36732632 · PMCID: PMC9928587

60. **CellBox: Interpretable Machine Learning for Perturbation Biology with Application to the Design of Cancer Combination Therapy**
Bo Yuan, Ciyue Shen, Augustin Luna, Anil Korkut, Debora S Marks, John Ingraham, Chris Sander
*Cell Systems* (2021-02) https://doi.org/ght4v6
DOI: 10.1016/j.cels.2020.11.013 · PMID: 33373583

61. **Effective gene expression prediction from sequence by integrating long-range interactions**
Žiga Avsec, Vikram Agarwal, Daniel Visentin, Joseph R Ledsam, Agnieszka Grabska-Barwinska, Kyle R Taylor, Yannis Assael, John Jumper, Pushmeet Kohli, David R Kelley
*Nature Methods* (2021-10) https://doi.org/gm2wv4
DOI: 10.1038/s41592-021-01252-x · PMID: 34608324 · PMCID: PMC8490152

62. **Transfer learning enables predictions in network biology**
Christina V Theodoris, Ling Xiao, Anant Chopra, Mark D Chaffin, Zeina R Al Sayed, Matthew C Hill, Helene Mantineo, Elizabeth M Brydon, Zexian Zeng, XShirley Liu, Patrick T Ellinor
*Nature* (2023-05-31) https://doi.org/gr9x63
DOI: 10.1038/s41586-023-06139-9 · PMID: 37258680

63. **scGPT: Towards Building a Foundation Model for Single-Cell Multi-omics Using Generative AI**
    Haotian Cui, Chloe Wang, Hassaan Maan, Kuan Pang, Fengning Luo, Bo Wang
    *Cold Spring Harbor Laboratory* (2023-05-01) https://doi.org/gshk6p
    DOI: 10.1101/2023.04.30.538439

64. **HyperAttention: Long-context Attention in Near-Linear Time**
    Insu Han, Rajesh Jayaram, Amin Karbasi, Vahab Mirrokni, David P Woodruff, Amir Zandieh
    *arXiv* (2023) https://doi.org/gtb9wc
    DOI: 10.48550/arxiv.2310.05869

65. **Stanford CRFM** https://crfm.stanford.edu/

66. **Assessing the limits of zero-shot foundation models in single-cell biology**
    Kasia Z Kedzierska, Lorin Crawford, Ava P Amini, Alex X Lu
    *Cold Spring Harbor Laboratory* (2023-10-17) https://doi.org/gszxk9
    DOI: 10.1101/2023.10.16.561085

67. **A Deep Dive into Single-Cell RNA Sequencing Foundation Models**
    Rebecca Boiarsky, Nalini Singh, Alejandro Buendia, Gad Getz, David Sontag
    *Cold Spring Harbor Laboratory* (2023-10-23) https://doi.org/gszxmb
    DOI: 10.1101/2023.10.19.563100

68. **Deep generative modeling for single-cell transcriptomics**
    Romain Lopez, Jeffrey Regier, Michael B Cole, Michael I Jordan, Nir Yosef
    *Nature Methods* (2018-11-30) https://doi.org/gfkw5z
    DOI: 10.1038/s41592-018-0229-2 · PMID: 30504886 · PMCID: PMC6289068

69. **Low-resource finetuning of foundation models beats state-of-the-art in histopathology**
    Benedikt Roth, Valentin Koch, Sophia J Wagner, Julia A Schnabel, Carsten Marr, Tingying Peng
    *arXiv* (2024) https://doi.org/gtdf95
    DOI: 10.48550/arxiv.2401.04720

70. **Are Emergent Abilities of Large Language Models a Mirage?**
    Rylan Schaeffer, Brando Miranda, Sanmi Koyejo
    (2023) https://openreview.net/forum?id=ITw9edRDlD

71. **ChatGPT broke the Turing test — the race is on for new ways to assess AI**
    Celeste Biever
    *Nature* (2023-07-25) https://doi.org/gskd92
    DOI: 10.1038/d41586-023-02361-7 · PMID: 37491395

72. **On the Opportunities and Risks of Foundation Models**
    Rishi Bommasani, Drew A Hudson, Ehsan Adeli, Russ Altman, Simran Arora, Sydney von Arx, Michael S Bernstein, Jeannette Bohg, Antoine Bosselut, Emma Brunskill, … Percy Liang
    *arXiv* (2021) https://doi.org/hw3v
    DOI: 10.48550/arxiv.2108.07258

73. **Designing an Interpretability-Based Model to Explain the Artificial Intelligence Algorithms in Healthcare**
    Mohammad Ennab, Hamid Mcheick
    *Diagnostics* (2022-06-26) https://doi.org/gtdf94
    DOI: 10.3390/diagnostics12071557 · PMID: 35885463 · PMCID: PMC9319389

74. **An enquiry concerning human understanding**
    David Hume, PF Millican

*Oxford University Press* (2007)
ISBN: 9780199211586

75. **Review of Dimension Reduction Methods**
    Salifu Nanga, Ahmed Tijani Bawah, Benjamin Ansah Acquaye, Mac-Issaka Billa, Francis Delali Baeta, Nii Afotey Odai, Samuel Kwaku Obeng, Ampem Darko Nsiah
    *Journal of Data Analysis and Information Processing* (2021) https://doi.org/gtb96v
    DOI: 10.4236/jdaip.2021.93013

76. **Why the simplest explanation isn't always the best**
    Eva L Dyer, Konrad Kording
    *Proceedings of the National Academy of Sciences* (2023-12-20) https://doi.org/gtbqms
    DOI: 10.1073/pnas.2319169120 · PMID: 38117857 · PMCID: PMC10756184

77. **The specious art of single-cell genomics**
    Tara Chari, Lior Pachter
    *PLOS Computational Biology* (2023-08-17) https://doi.org/gtb96t
    DOI: 10.1371/journal.pcbi.1011288 · PMID: 37590228 · PMCID: PMC10434946

78. **The Uncanny Failure of A.I.-Generated Hands**
    Kyle Chayka
    *The New Yorker* (2023-03-10) https://www.newyorker.com/culture/rabbit-holes/the-uncanny-failures-of-ai-generated-hands

79. **Identifying Representations for Intervention Extrapolation**
    Sorawit Saengkyongam, Elan Rosenfeld, Pradeep Ravikumar, Niklas Pfister, Jonas Peters
    *arXiv* (2023) https://doi.org/gtb97m
    DOI: 10.48550/arxiv.2310.04295

80. **The Causal-Neural Connection: Expressiveness, Learnability, and Inference**
    Kevin Xia, Kai-Zhan Lee, Yoshua Bengio, Elias Bareinboim
    *arXiv* (2021) https://doi.org/grw6m7
    DOI: 10.48550/arxiv.2107.00793

81. **Exploring the Latent Space of Autoencoders with Interventional Assays**
    Felix Leeb, Stefan Bauer, Michel Besserve, Bernhard Schölkopf
    *arXiv* (2021) https://doi.org/gtb96x
    DOI: 10.48550/arxiv.2106.16091

82. **Improving black-box optimization in VAE latent space using decoder uncertainty**
    Pascal Notin, José Miguel Hernández-Lobato, Yarin Gal
    *arXiv* (2021) https://doi.org/gtb962
    DOI: 10.48550/arxiv.2107.00096

83. **On the Number of Experiments Sufficient and in the Worst Case Necessary to Identify All Causal Relations Among N Variables**
    Frederick Eberhardt, Clark Glymour, Richard Scheines
    *arXiv* (2012) https://doi.org/gtb997
    DOI: 10.48550/arxiv.1207.1389

84. **Learning Neural Causal Models from Unknown Interventions**
    Nan Rosemary Ke, Olexa Bilaniuk, Anirudh Goyal, Stefan Bauer, Hugo Larochelle, Bernhard Schölkopf, Michael C Mozer, Chris Pal, Yoshua Bengio
    *arXiv* (2019) https://doi.org/grw6nc
    DOI: 10.48550/arxiv.1910.01075

85. **DiscoBAX - Discovery of optimal intervention sets in genomic experiment design**
Clare Lyle, Arash Mehrjou, Pascal Notin, Andrew Jesson, Stefan Bauer, Yarin Gal, Patrick Schwab
(2023) https://openreview.net/forum?id=mBkUeW8rpD6

86. **Interventions, Where and How? Experimental Design for Causal Models at Scale**
Panagiotis Tigas, Yashas Annadani, Andrew Jesson, Bernhard Schölkopf, Yarin Gal, Stefan Bauer
*arXiv* (2022) https://doi.org/gtcfvk
DOI: 10.48550/arxiv.2203.02016

87. **CausalBench: A Large-scale Benchmark for Network Inference from Single-cell Perturbation Data**
Mathieu Chevalley, Yusuf Roohani, Arash Mehrjou, Jure Leskovec, Patrick Schwab
*arXiv* (2022) https://doi.org/gtcbbc
DOI: 10.48550/arxiv.2210.17283